\documentclass[12pt,reqno]{article}
\usepackage{chet}
\usepackage{booktabs}
\usepackage{amsmath,hyperref,amssymb}
\pdfoutput=1
\usepackage{graphicx,url}
\usepackage{color}
\usepackage{setspace}
 
\usepackage{epsfig}

\def\Or[#1]{{\text{O}}\left({#1}\right)}
\def\dotl[#1,#2]{\left\langle #1, #2 \right\rangle}
\def\dotlb[#1,#2]{[ #1, #2 ]}
\def\dotp[#1,#2]{(#1) \cdot (#2)}
\def\aff[#1,#2]{\hat{#1}(#2)}
\def\n4sym{{\cal N}=4 SYM}
\def\>{\rangle}
\def\<{\langle}
\def\weight[#1,#2,#3]{\{(#1),#2,#3\}}
\def\ads[#1]{$\text{AdS}_{#1}$}

\newcommand{\ba}{\begin{eqnarray}}
\newcommand{\ea}{\end{eqnarray}}
\newcommand{\be}{\begin{eqnarray}}
\newcommand{\ee}{\end{eqnarray}}
\newcommand{\bq}{\begin{equation}}
\newcommand{\eq}{\end{equation}}
\newcommand{\benn}{\begin{equation*}}
\newcommand{\eenn}{\end{equation*}}
\newcommand{\bi}{\begin{itemize}}  
\newcommand{\ei}{\end{itemize}}

\newcommand{\CO}{{\cal O}}

\newcommand{\CV}{{\cal V}}
\newcommand{\nn}{\nonumber}

\newcommand\oo\infty
\newcommand\s\sigma
\newcommand\de\delta
\newcommand\De\Delta

\newcommand\f\phi
\newcommand\g\gamma
\newcommand\x\times

\begin{document}

\title{A Quantum Correction To Chaos}
\author{A. Liam Fitzpatrick$^1$, Jared Kaplan$^2$}
\affiliation{
{\it $^1$ Dept.\ of Physics, Boston University, Boston, MA 02215} \\ 
{\it $^2$ Dept. of Physics and Astronomy, Johns Hopkins University, Baltimore, MD 21218} \\
}
\abstract{
We use results on Virasoro conformal blocks to study chaotic dynamics in CFT$_2$ at large central charge $c$.   The Lyapunov exponent $\lambda_L$, which is a diagnostic for the early onset of chaos, receives $1/c$ corrections that may be interpreted as $\lambda_L = \frac{2 \pi}{\beta} \left( 1 + \frac{12}{c} \right)$.  However, out of time order correlators receive other equally important $1/c$ suppressed contributions that do not have such a simple interpretation.  We revisit the proof of a bound 
on $\lambda_L$ that emerges at large $c$, focusing on CFT$_2$ and explaining why our results do not conflict with the analysis leading to the bound.  We also comment on relationships between chaos, scattering, causality,  and bulk locality.
}
 
\maketitle

\tableofcontents

\flushbottom

\section{Introduction}

The commutation relations of local operators are a canonical avatar of causality.  At time-like separation one operator can have a causal affect on another, leading to non-vanishing commutators.  However,  such commutators are not necessarily large:
 the death of a single butterfly is unlikely to affect the outcome of a presidential election.  
It is only in some theories that chaos is generic and large commutators of local operators are the rule rather than the exception\cite{georges2001quantum,d2015quantum,bradbury1952sound,kitaev2014hidden,Shenker:2013pqa, Hosur:2015ylk, Gur-Ari:2015rcq, Berenstein:2015yxu, Stanford:2015owe}.  Recently, it has been conjectured that theories dual to gravity are in a sense as chaotic as possible \cite{Maldacena:2015waa,Shenker:2014cwa} and the rate of growth of commutators of  local operators saturate a general upper bound on all theories.

Virasoro conformal blocks are a useful tool for studying this conjecture in greater depth, since they encapsulate the effect of multi-graviton exchanges in AdS$_3$.  The fact that all graviton excitations are related to the vacuum by Virasoro symmetry renders their effect computable in principle, and a number of methods allow them to be computed in practice in various expansions.  Most relevantly for gravity, their large $c$ expansion is dual to a large $m_{\rm pl}$ limit \cite{Fitzpatrick:2014vua, Fitzpatrick:2015zha, Fitzpatrick:2015foa, Asplund:2014coa, Hijano:2015rla, Alkalaev:2015wia, Hijano:2015qja,Fitzpatrick:2015dlt, Hongbin,HartmanLargeC,Brown:1986nw}.  In fact, as we will review, it is only in such a large  $c$ limit that the bound of MSS \cite{Maldacena:2015waa} takes a sharp form, 
\be
| \dot{F}(t)|  \le \frac{2 \pi}{\beta}  \left( 1 - F(t) + \epsilon \right),
\label{eq:MSSBound}
\ee
where $F$ is a certain out of time ordered correlator of local operators, $\beta^{-1}$ is the temperature, and $\epsilon$ is a small correction containing non-universal terms that vanishes at $c\rightarrow \infty$.\footnote{The correction $\epsilon$ is in a sense time-dependent, but it is convenient to define it to be time-independent by taking it to be the supremum of the correction over the time region of interest.}  The bound holds for times $t > t_d$ where $t_d \sim \beta$ is a dissipation time scale.  We will be interested in times $t < t_*$, where $t_* \propto \log(c) \gg t_d$ is the scrambling time defined by $| 1- F(t_*)| \sim 1$.

At leading order in $1/c$, the out of time order correlator can be computed in gravity duals \cite{Shenker:2014cwa},  and has been found to saturate the bound in a specific regime.   In any regime $t \in (t_d, t_f)$ where $1-F(t)$ grows exponentially we may write
\be
1-F(t) = A e^{\lambda_L t}, \qquad (t_1 \le t \le t_2),
\label{eq:CexpForm}
\ee
the bound (\ref{eq:MSSBound}) is conveniently stated as a bound on the exponent $\lambda_L$,
\be
\lambda_L \le \frac{2\pi}{\beta} \left( 1 + \frac{\epsilon}{A} e^{- \lambda t_2} \right).
\label{eq:MSSBound2}
\ee
The size of $\epsilon$ depends on the regime of interest; we will see in section \ref{sec:ReviewBound} that for $t\gtrsim \beta$, one has $\epsilon \sim c^{-1}$, whereas for $t \gtrsim \frac{\beta}{2\pi} \log c$ one has $\epsilon \sim c^{-3}$.   In the main regime of interest, $\lambda_L$ is called the Lyapunov exponent.  All of $A , \epsilon $, and $t_2 \sim \frac{\beta}{2\pi} \log c$ depend on $c$, but at $c\rightarrow \infty$ their combination vanishes, and the leading large $c$ value $\lambda_L = \frac{2\pi}{\beta}$ in gravity theories saturates the resulting bound.  

  The main goal of this paper is to explore the purely gravitational quantum correction to the growth of $1-F(t)$, which we will study by analyzing the Virasoro vacuum block at higher orders in $1/c$.  Using recent results \cite{Fitzpatrick:2014vua, Fitzpatrick:2015zha, Fitzpatrick:2015foa, Asplund:2014coa, Hijano:2015rla, Hijano:2015qja,Fitzpatrick:2015dlt, Hongbin,Headrick}, we directly compute $F(t)$ and find in the regime $\beta \lesssim t \sim \frac{\beta}{2\pi} \log c$ that there are corrections of the form
  \be
1-  F(t) = A e^{\frac{2 \pi}{\beta} t} \left( 1 + \frac{2 \pi}{\beta} \frac{\delta}{c} t + \dots \right),
\label{eq:corrections}
  \ee
  where $\delta$ is a pure number independent of the conformal weights of the operators, say $W$ and $V$, in the correlator $F(t)$. At leading order in $1/c$, we find
  \be
  \delta = 12 + \CO \left( \frac{1}{c} \right).
  \label{eq:anomdimdelta}
  \ee
 The form of the correction looks like a perturbative correction to the exponent 
  \be
  \lambda_L \approx \frac{2 \pi}{\beta} \left(1 + \frac{12}{c} \right).
  \ee
  In forthcoming work \cite{Hongbin}, we will explore higher order corrections in $1/c$ in more detail,  and prove that higher order corrections in $1/c$ exponentiate.

      Since the real part of $\delta$ is positive, this naively looks like a violation of (\ref{eq:MSSBound2}).  However, one has to be more precise both about the extra terms $\dots$ in (\ref{eq:corrections}) and about the small non-universal correction factor $\epsilon$. In our review of the proof in \cite{Maldacena:2015waa}, we will emphasize that it is consistent with $\delta$ having either positive or negative sign at early $t \gtrsim t_d$, the dissipation time.  In fact, the positive anomalous dimension (\ref{eq:anomdimdelta}) will be allowed for different reasons in different regimes: for $t \sim \beta$, we will see that the factor $\epsilon$ is sufficiently large to allow for the anomalous dimension (\ref{eq:anomdimdelta}), whereas for $t\sim \frac{2 \pi}{\beta} \log c$, the correction terms $\dots$ contain additional contributions to the growth of $1-F$ that are not of the form of an anomalous dimension and are parametrically larger, bringing the rate of growth below the bound.
  
  \begin{figure}[t!]
\begin{center}
\includegraphics[width=0.7\textwidth]{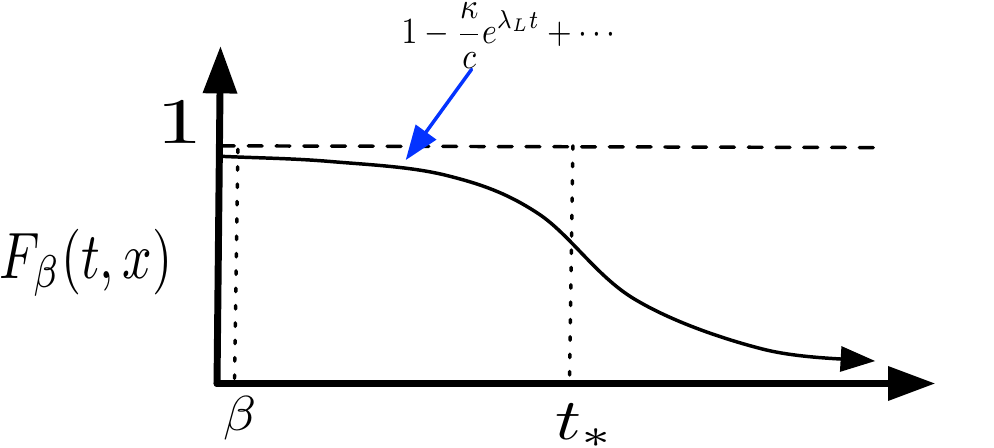}
\caption{ This figure depicts the various regimes of the out of time order correlator $F_\beta$ of equation (\ref{eq:OutOfOrderCorr}).  At early times $t \sim \beta$ we expect $1-F_\beta \lesssim \frac{1}{c}$; this quantity subsequently grows exponentially with Lyapunov exponent $\lambda_L$ until a scrambling time of order $t_* \approx \frac{\beta}{2 \pi} \log c$. }
 \label{fig:ChaoticBehaviorLyapunov} 
\end{center}
\end{figure}

The outline of the paper is as follows.  In section \ref{sec:ReviewBound}, we review the recent proof in \cite{Maldacena:2015waa}, framing it in language closer to that of the conformal bootstrap, and focusing on the potential for $1/c$ corrections in CFT$_2$.    In section \ref{sec:QuantumCorrections}, we present our computation of the $1/c$ correction from gravity. 
 In section \ref{sec:AdS}, we discuss the regime in which the Virasoro vacuum block is expected to dominate the answer, and the relation to various regimes of bulk physics in the gravity dual.  In particular, we explore connections between bounds on chaos and bounds from causality and unitarity in both flat space scattering and AdS/CFT.     Finally, in section \ref{sec:discussion}, we discuss possible implications and  future directions.

\section{Lyapunov Exponents and Bounds}
\label{sec:ReviewBound}

We will be studying thermal correlators with various time orderings in order to define the Lyapunov exponent $\lambda_L$, a measure of the onset of chaos.  Throughout the paper $V$ and $W$ will be local primary operators in a CFT$_2$.  Much of the discussion will be a review of the setup \cite{Roberts:2014ifa, Maldacena:2015waa} and various analytic continuations and coordinate choices \cite{Hartman:2015lfa, Maldacena:2015iua}.  But we will  streamline the derivation of the bound \cite{Maldacena:2015waa} by specializing to the case of CFTs, and we will be very precise about potential $1/c$ corrections, which we will  compute in subsequent sections.

\subsection{Review of the Kinematics}

We would like to study a finite temperature correlator probing chaos in CFTs.  Perhaps the most natural observable would be a squared commutator, but 
to connect more closely with \cite{Maldacena:2015waa} we will study a related 
out of time order correlator of two local operators $V$ and $W$ in a thermal background:
\be
F_\beta(x_i) &=& \< V(x_1) W(x_3) V(x_2)   W(x_4)\>_\beta.
\label{eq:OutOfOrderCorr}
\ee
In the Euclidean regime, all local operators commute, but in the Lorentzian regime the operator ordering of equation \ref{eq:OutOfOrderCorr} is important.  Any Lorentzian operator ordering can be obtained from the Euclidean correlator by analytically continuing through a sequence of branch cuts.

 \begin{figure}
 \begin{center}
 \includegraphics[width=0.85\textwidth]{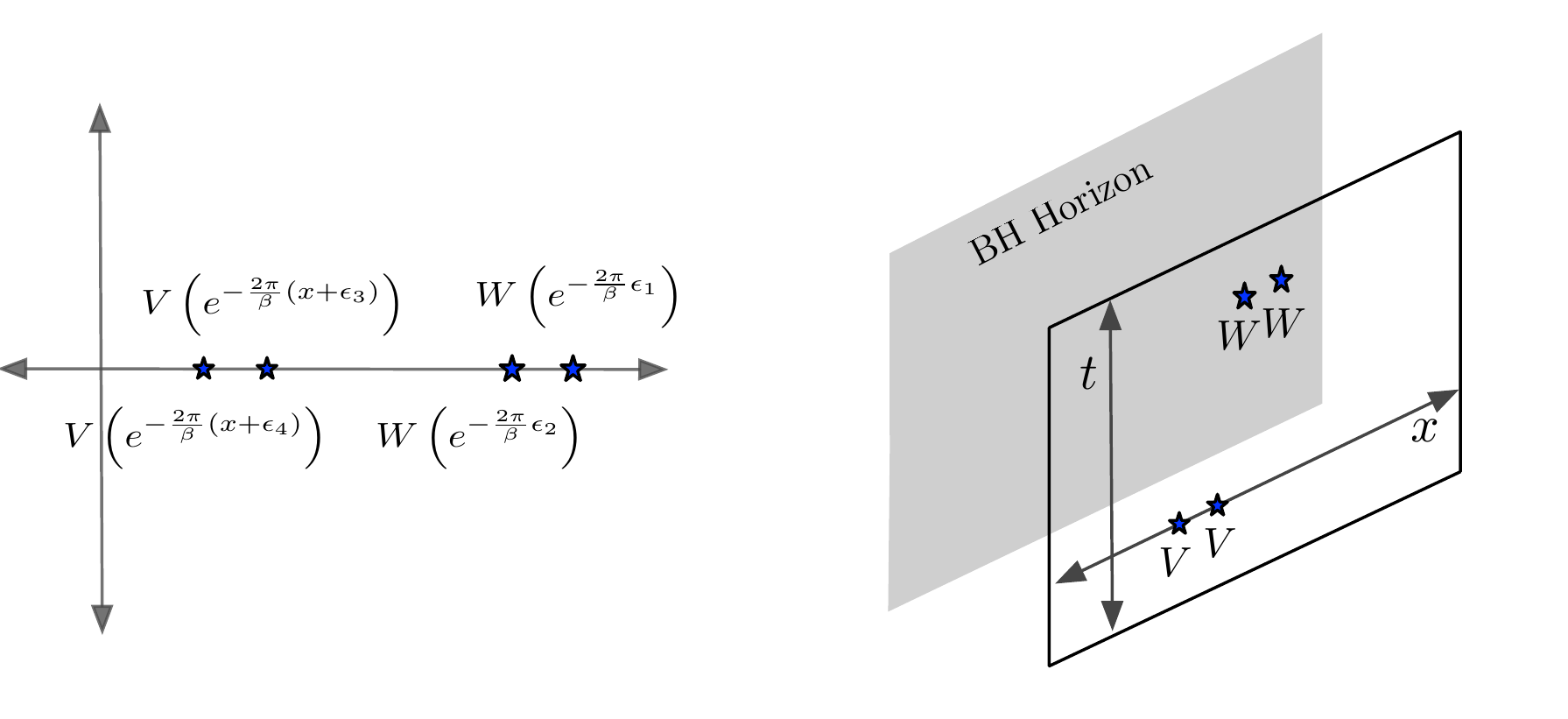}
 \end{center}
\caption{ The figure on the left shows a choice of locations for $V$ and $W$ \cite{Roberts:2014ifa} in the Euclidean plane at Lorentzian $t=0$.  The distance between the two $V$ and the two $W$ is small, regulated by $\epsilon_i - \epsilon_j$.  On the right we have a cartoon of the configuration at finite $x, t$ from the perspective of AdS$_3$/CFT$_2$.  This configuration has the physically intuitive advantage that all operators are on the same side of a single black brane in the Poincar\'e patch.}
 \label{fig:RobertsStanfordCoords}
 \end{figure}

We will mainly be interested in 2d CFTs, where we have two advantages:  we can connect correlators in the plane to finite temperature correlators via a conformal transformation, and we can use the Virasoro algebra to uniquely determine certain universal contributions to any correlator.  Note that the finite temperature $T = 1/\beta$ breaks both conformal and Lorentz invariance.  The temperature is the unique dimensionful quantity in our setup, so its numerical value is irrelevant as long as $T \neq 0$.  

\begin{figure}
 \begin{center}
 \includegraphics[width=0.95\textwidth]{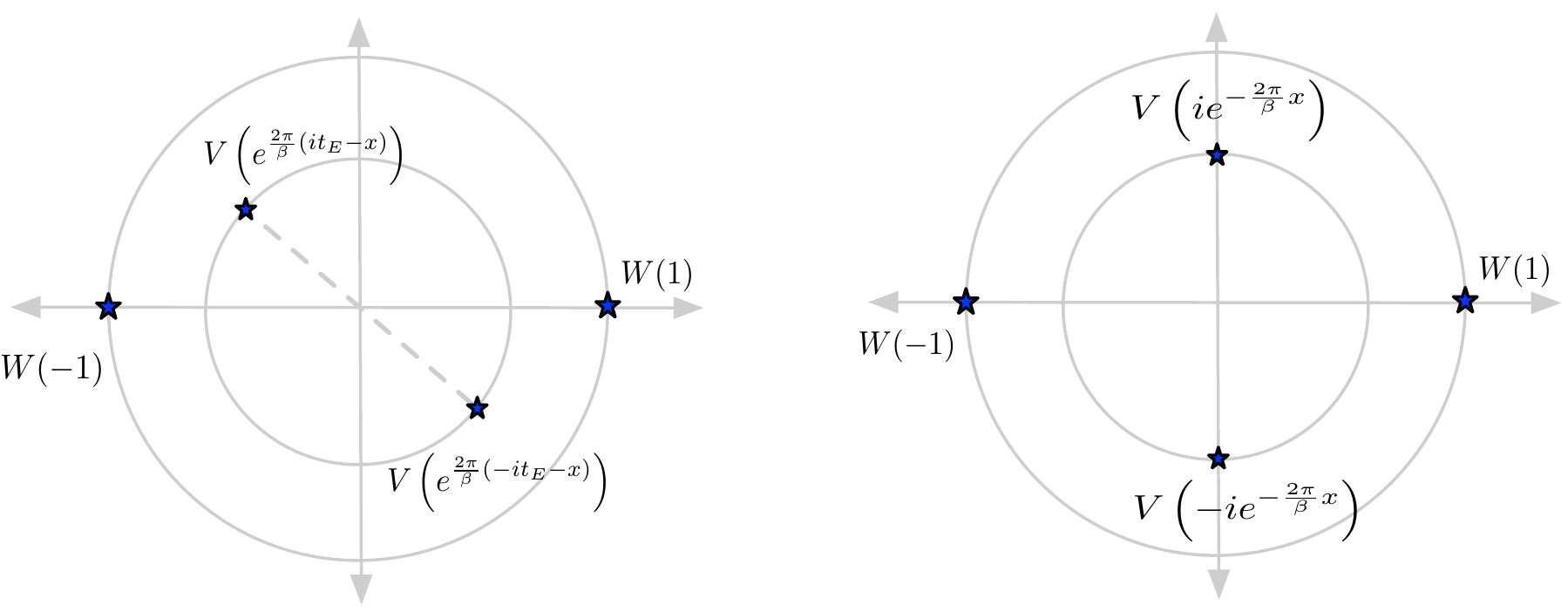}
 \end{center}
\caption{This figure shows the locations of the operators in the correlator $F_\beta(x_i)$.  We work at finite temperature, so the Euclidean time $t_E$ is an angular variable with period $\beta$.  On the left we have a general Euclidean configuration in the $\rho$ coordinates of equation \ref{eq:rhodefinition}, while the on the right is the Lorentzian $t= 0$ slice of the configuration relevant to the study of chaos, with operators spaced out over $\beta/4$ intervals in Euclidean time.}
 \label{fig:OpLocation}
 \end{figure}

From the AdS perspective, chaos arises from scattering near the horizon of a black hole.  It is possible to make a choice of coordinates \cite{Roberts:2014ifa} for $F_\beta(x_i)$ that emphasizes this aspect of the physics, as suggested by Figure \ref{fig:RobertsStanfordCoords}.  However, this choice introduces additional parameters to define the distance between the $VV$ and $WW$ operator pairs, and so we will instead use a technically more convenient choice\footnote{For another recent discussion see \cite{Maldacena:2015iua}.} of coordinates \cite{Maldacena:2015waa,Shenker:2014cwa} where the $VV$ and $WW$ separations are proportional to $\beta$, as displayed in the Euclidean plane in Figure \ref{fig:OpLocation}.   Note that unlike in conventional radial quantization, the Euclidean time $t_E$ plays the role of an angular coordinate with thermal period $\beta$.    This choice has a natural correspondence with the $\rho$ coordinates  \cite{Pappadopulo:2012jk}, which provide the largest possible radius of convergence for the conformal block decomposition  in a general spacetime dimension.   The $\rho$ coordinates map the $z$ coordinates on the plane\footnote{where the operators are located at $\< \CO(\infty) \CO(1) \CO(z) \CO(0) \>$} to the unit disk:
 \be
 \rho(z) &=& \frac{z}{(1+\sqrt{1-z})^2}.
\label{eq:rhodefinition}
 \ee
In other words, the operators are located at $\pm 1$ and $\pm \rho$, as shown in figure \ref{fig:OpLocation}.
The region of convergence of the $VV$ OPE is $|\rho|<1$, corresponding to the entire cut $z$-plane.

Thus we will follow the authors of \cite{Maldacena:2015waa,Shenker:2014cwa}, who considered the four-point function $F_\beta(x_i)$ with the operators spread out around the thermal cycle, at Euclidean times $\tau_1 = \beta/4, \tau_2 = 3\beta/4$ and $\tau_3 = 0, \tau_4=\beta/2$.
 At finite Lorentzian time $t$ and spatial separation $x$, we take the coordinates on the plane to be $x_1, \bar{x}_1 = \pm i e^{2 \pi \beta^{-1} (\pm t-x)},~ x_2, \bar{x}_2 =\mp i e^{ 2\pi \beta^{-1}(\pm t-x)},~x_3, \bar{x}_3 = 1,~ x_4, \bar{x}_4 = -1$.  As is evident from Figure \ref{fig:OpLocation}, the $VV$ and $WW$ OPEs only converge for $x > 0$; the regime of large $x$ is an OPE limit, and at large Lorentzian $t$ it is a lightcone OPE limit \cite{Fitzpatrick:2012yx, KomargodskiZhiboedov}.  The conformal transformation from the plane to the thermal cylinder also requires us to multiply the correlator in the plane by a simple overall factor \cite{Roberts:2014ifa}.
 
 We can view the correlator $F_\beta(x_i)$ as the overlap between the states $V(x_2) W(x_4)| 0\>$ and $W^\dagger(x_3) V^\dagger(x_1)|0\>$, with the physical intuition that at large times chaotic dynamics should make these states very different.
We will not be focused on the asymptotically large $t$ behavior of equation (\ref{eq:OutOfOrderCorr}), but in its behavior at intermediate times $\beta \ll t \ll t_*$, where $t_*$ is referred to as the scrambling time.  In other words, we are interested in the initial development of chaos, and the analysis can only apply to theories with $t_* \gg \beta$.  For our purposes $t_* = \frac{\beta}{2 \pi} \log c$ where $c$ is the central charge of a CFT.

Normalizing to the disconnected correlator, we therefore expect that in the regime $\beta \ll t \ll t_*$ with $t > x$ we have
\be
\frac{F_\beta(x_i)}{\<V(x_1) V(x_2)\>\<W(x_3)W(x_4)\>}
& \approx &  1 - \frac{\kappa}{c} e^{\lambda_L(x) t}  + \cdots
\label{eq:Fapprox}
\ee
where $\kappa$ is a $c$-independent constant that depends on the kinematics and the properties of $V$ and $W$.  The exponential behavior, which is expected if the divergence of nearby trajectories is proportional to their separation in phase space, defines the Lyapunov exponent $\lambda_L(x)$.
We have emphasized that the exponent may depend on the spatial separation $x$ between the pairs of operators $VV$ and $WW$.  Stringy corrections \cite{Shenker:2014cwa} provide an explicit and physically interesting example where the Lyapunov exponent depends on $x$.    

We will see in section \ref{sec:QuantumCorrections} that at the higher order in $1/c$, there are corrections to equation (\ref{eq:Fapprox}) that can be naturally interpreted as a $\frac{1}{c}$ shift in $\lambda_L$.  However, there are also other corrections, such as a term $\frac{1}{c^2} e^{\lambda' t}$, which are of equal or greater numerical importance in the regime $\beta \ll t \ll t_*$.

\subsection{A Bound on the Lyapunov Exponent and $\frac{1}{c}$ Corrections}
\label{sec:Bound}

Now we will discuss the correlator (\ref{eq:OutOfOrderCorr}) and review the bound on its growth.
As pictured in Figure \ref{fig:OpLocation}, we use coordinates
 \be
 \rho = i e^{ \frac{2 \pi (t-x)}{\beta}}, \qquad \bar{\rho} = -i e^{\frac{2\pi(-t-x)}{\beta}}.
 \ee
At $t=0$ and $x$ large, the two insertions of $V$ come very close to each other in the Euclidean regime, and so the correlator is controlled by the Euclidean OPE limit; in this regime, $z \approx 4 i e^{-\frac{2 \pi x}{\beta}}$.   We will mainly consider the regime of large $x$ and positive $t$, so that $\bar{\rho}$ stays very small and well within the radius of convergence of the sum over blocks.  This is a lightcone OPE limit \cite{Fitzpatrick:2012yx, KomargodskiZhiboedov}.  What this means in practice is that contributions from the lowest-twist operator in the sum over blocks will dominate the correlator.\footnote{Conformal blocks began at order $\bar{\rho}^{~\tau/2}$ in a small $\bar{\rho}$ expansion, where $\tau = \Delta-\ell = 2\bar{h}$ is the twist of the primary operator.}

\begin{figure}[t!]
\begin{center}
\includegraphics[width=0.45\textwidth]{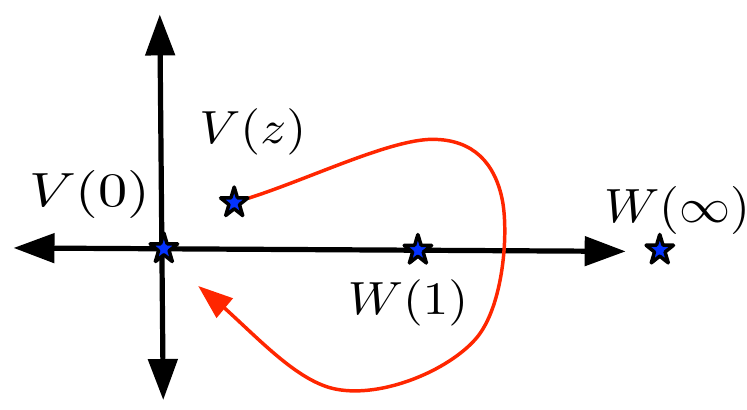}
\caption{ CFT 4-point correlators can have branch cuts between their OPE singularities at $0, 1,$ and $\infty$. This figure shows the analytic continuation in the $z$ plane necessary to obtain the out of time order correlator of equation (\ref{eq:OutOfOrderCorr}).  While $z$ pases through the branch cut extending from $1$ to $\infty$, the $\bar z$ variable remains on the original sheet.  }
 \label{fig:AnalyticContinuationChaos} 
\end{center}
\end{figure}

As we increase $t$ to values larger than $x$, the $\rho$ variable passes outside the radius of convergence of the global blocks, and so one potentially becomes sensitive to all operators in the conformal block expansion.  If we track $z$ in terms of $t$,
  \be
  z     &=& 1-\left( \frac{1-i e^{\frac{2 \pi (t-x)}{\beta}}}{1+ i e^{\frac{2 \pi (t-x)}{\beta}}} \right)^2,
  \ee
  we see that $z$ starts out near $0$ at small $t$, passes around the insertion of $W$ at $1$, and returns to $z \sim 0$, as depicted in Figure \ref{fig:AnalyticContinuationChaos}. Due to the insertion of $W$, there is a branch cut extending from 1 to $\infty$ in the $z$ plane, and we have to make a choice about whether to pass through it onto the second sheet of the correlator, or to deform the path in order to avoid the cut.  Passing through the cut corresponds to exchanging the order of operators from the radial OPE ordering $\< W(\infty) W(1) V(z) V(0)\>$, and so in order to obtain the correlator with operators ordered \cite{Hartman:2015lfa} as in (\ref{eq:OutOfOrderCorr}) at large $t$, $z$ approaches 0 on the second sheet of the correlator. 
 
Next, a crucial part of the proof is that for ${\rm Re}(t)$ sufficiently large, the correlator is analytic for $|{\rm Im}(t)| \le \frac{\beta}{4}$ and bounded  for ${\rm Im}(t) = \pm \frac{\beta}{4}$.  We will review the proof in the language  of the recently derived CFT causality constraints  \cite{Hartman:2015lfa}. At  large $t$, one has $(z, \bar{z}) \approx (\frac{4}{\rho}, 4 \bar{\rho})$, which approaches the origin with $z$ on the second sheet and $\bar{z}$ on the first sheet.  In the notation of \cite{Hartman:2015lfa}, this is
\be
z = \sigma, \qquad \bar{z} = \eta \sigma,
\ee
with 
\be
\label{eq:Sigmaoft}
\eta \approx \rho \bar{\rho} = e^{ -\frac{4 \pi x}{\beta}} , \qquad \sigma \approx \frac{4}{\rho} = -4i e^{\frac{2\pi (x-t)}{\beta}}.
\ee
The commutator $[ V(x_1), V(x_2)]$ vanishes outside the lightcone in the background state $|W\> = W(0) | 0 \>$ created by the operator $W$ if and only if $F_\beta$ is an analytic function of $\sigma$ for ${\rm Im}(\sigma) \le 0$ in a neighborhood of $\sigma \sim 0$ \cite{Hartman:2015lfa}. This is the same criterion as $| {\rm Im}(t) | \le \frac{\beta}{4}$. (As we will discuss shortly in detail, a similar analytic continuation is used to study 2-to-2 scattering in AdS/CFT \cite{GGP}).    Furthermore, while the $V(z)V(0)$ OPE does not converge on the second sheet at $z \sim 0$, the $V(z) W(1)$ OPE does converge in this region on both the first and the second sheet.  It is a sum over positive terms on the first sheet, and analytically continuing to the second sheet $(1-z)\rightarrow e^{2 \pi i}(1-z)$ introduces phases that can only {\it decrease} the magnitude of the correlator.  Thus, at $z \sim 0$ on the second sheet, the correlator is still bounded by its value on the first sheet, where the $VV$ OPE does converge.

This is sufficient to prove that a bound of the form (\ref{eq:MSSBound}) mentioned in the introduction exists for some $\epsilon$.  
Consider the correlator $F_\beta$ normalized by the disconnected piece.
It is bounded by its behavior on the first (Euclidean) sheet, where we can analyze it using the $VV$ OPE channel
\be
\left|\frac{F_\beta(x_i)}{ \< V(x_1) V(x_2)\> \< W(x_3) W(x_4)\>}\right| \le \sum_{(h_\alpha,\bar{h}_\alpha)} c_{VV \alpha} c_{WW \alpha} g_{h_\alpha, \bar{h}_\alpha}(z,\bar{z}),
\label{eq:FBlockDecomp}
\ee
where $g_{h,\bar{h}}(z,\bar{z})$ are the global conformal blocks
\be
g_{h,\bar{h}}(z,\bar{z}) &=& g_h(z) g_{\bar{h}}(\bar{z}), \nn\\
g_h(z) &=& z^h {}_2F_1(h,h,2h,z).
\ee
 In CFT$_2$ we can use either the $SO(1,3)$  global conformal blocks or the full Virasoro conformal blocks; however the latter are only known in closed form in a certain special limits, so we have used global blocks for concreteness.  The $c_{VV \alpha}, c_{W W \alpha}$ are OPE coefficients proportional to $\< VV \CO_{\alpha}\>$ and $\<WW\CO_{\alpha}\>$. 
 
Now, by taking $x$  large, we can make $\eta$ and $\bar{z}$ small, and therefore we can suppress terms with $\bar{h}_\alpha>0$.  This means that the zero-twist states, i.e. $\bar{h}=0$, dominate in this regime.  In a completely general 2d conformal theory, the zero-twist sector can include higher-spin current and be very complicated.  However, in theories whose gravity duals are purely GR at low energies, or GR plus massive fields, the only twist-zero sector states are multi-stress tensor products, which make up  the Virasoro vacuum conformal block.  Therefore, their contribution is in principle known and can be easily estimated. By taking $t-x$ large as well, we can also make $\sigma$ small.  However, we cannot take this to be arbitrarily small, since we will mainly be interested in the regime where we are starting to approach the scrambling time $t_* \approx \frac{\beta}{2 \pi} \log(c) +x$ but are still much earlier than it, so that $\frac{1}{c} e^{\frac{2 \pi (t-x)}{\beta}} \ll 1$.  To be precise, we want to take the limit where $|\sigma| c$ is held constant and large as $c\rightarrow \infty$. 

The upshot is that $\frac{F_\beta}{\< V V\>\<W W\>}$ is parametrically bounded by the leading contributions from the vacuum Virasoro block and contributions with minimal twist $\tau_m$ on the first sheet.  The former of these scale like $\frac{h_L h_H}{c} \sigma^2$, and the latter scale like $ \sigma^{\Delta_m} \eta^{\tau_m}$, where $\Delta_m, \tau_m$ are the dimension and twist of the minimal twist operator (after the vacuum Virasoro block):
\be
\left| \frac{F_\beta(x_i)}{ \< V(x_1) V(x_2)\> \< W(x_3)W(x_4)\>} \right|  &\lesssim& 1+  2\frac{h_W h_V}{c} |\sigma|^2  + A_m |\sigma|^{\Delta_m} \eta^{\tau_m} \nn\\
&=& 1+ 2 \frac{h_W h_V}{c^3} |\sigma c|^2 + e^{- \frac{4 \pi \tau_m x}{\beta}}  \frac{A_m}{c^{\Delta_m}}|\sigma c|^{\Delta_m} ,
\label{eq:boundsummary}
\ee
where $A_m$ is an unknown theory-dependent prefactor.  In $d>2$, the first term would be absent, since in a unitary theory, operators with twist below $\frac{d-2}{2}$ are forbidden and thus all contributions, even conserved currents, are suppressed by a $e^{- \frac{4 \pi \tau_m x}{\beta}}$ factor. Thus, $\frac{F_\beta}{\< V V\>\<W W\>}$ is bounded and analytic in the region ${\rm Im}(\sigma)<0$. Taking $f(t)= \frac{F_\beta}{\< V V\>\<W W\>}$ and letting $f_m$ be the maximum value of the RHS of (\ref{eq:boundsummary}) for times later than some optimally chosen time $t_0$ (i.e., for $\sigma$ less than some value), an elementary complex analysis argument  \cite{Maldacena:2015waa} then implies that
\be
\left| \frac{df}{dt} \right| \le \frac{2 \pi}{\beta} \left( f_m -|f|\right) \frac{\left(1 + \frac{|f|}{f_m}\right) }{2} \left( 1 + \sinh^{-2} \left( \frac{2 \pi t}{\beta} \right) \right) .
\ee
 
A crucial point in the above inequality is that the irreducible non-universal term $\epsilon$ in (\ref{eq:MSSBound}) arises from the terms on the RHS of (\ref{eq:boundsummary}).  Recalling $\sigma \approx  -4i e^{\frac{2\pi (x-t)}{\beta}}$, this means that at times $t \propto t_*$ the error $\epsilon \sim \CO\left( \frac{h_W h_V}{c^3} \right)$, but at times $t \gtrsim \beta$ we have $\epsilon \sim \CO\left( \frac{h_W h_V}{c} \right)$.  In other words, if we wish to constrain $f(t)$ for all times $t > t_0$ with $t_0 \sim \beta$ then we can only bound $| f'(t) |$ up to corrections of order $1/c$.  But if we are specifically interested in the more restricted regime where $t_0$ scales with $t_* = \frac{\beta}{2 \pi} \log c$, then we can obtain a parametrically stronger bound on chaos.

\begin{center}
\begin{tabular}{c}\hline\phantom{ssssssssss}\end{tabular}
\end{center}

Finally, let us review the qualitative behavior of the conformal block decomposition in the $V V$ OPE channel on the second sheet, and in particular its behavior in different regimes of the spatial separation $x$.   Passing to the second sheet via  term-by-term analytic continuation of the conformal blocks in (\ref{eq:FBlockDecomp}), one obtains \cite{Roberts:2014ifa}
\begin{equation}\label{eq:FfromGlobalBlocks}
\frac{F_\beta(x_i)}{ \< V(x_1) V(x_2)\> \< W(x_3)W(x_4)\>} \approx \sum_{(h_\alpha,\bar{h}_\alpha)} \tilde{c}_{V V \alpha} \tilde{c}_{W W \alpha}  e^{(h_\alpha-\bar h_\alpha - 1)t - (h_\alpha + \bar h_\alpha -1)x} \left( 1 + O \! \left(e^{-x}, e^{-t} \right) \right),
\end{equation}
where we note that $h, \bar h \geq 0$ in unitary theories, and we identify the spin $\ell = h- \bar h$ and the total dimension $\Delta = h + \bar h$.  The $\tilde{c}_{V V \alpha} \tilde{c}_{W W \alpha} $ have been rescaled to absorb some $h$-dependent coefficients.  The conformal block decomposition of equation (\ref{eq:FfromGlobalBlocks}) will not in general converge.  In the presence of an unbounded sum over spins $\ell$, it appears that the Lyapunov exponent cannot be extracted until after performing the sum.  We will discuss  the infinite sum over ubiquitous `double-trace' operators \cite{Fitzpatrick:2012yx, KomargodskiZhiboedov,Polchinski:2002jw,JP} in section \ref{sec:DoubleTraceContributions}.

We see from the structure of equation (\ref{eq:FfromGlobalBlocks}) that the contributions of large dimension operators will be suppressed at large $x$, leading to an $x$-dependent Lyapunov exponent.  The limit of large $x$ and $t$ is a lightcone OPE limit \cite{Fitzpatrick:2012yx, KomargodskiZhiboedov} for the CFT correlator; this limit suppresses contributions from large twist.  For example, AdS string states are massive, and therefore correspond to large twist operators in the CFT, which do not affect the $\lambda_L(x)$ at large $x$  \cite{Shenker:2014cwa}. 
 In circumstances where only a single conformal block dominates at large $c$, or when we can sum all relevant contributions, we can make predictions about $\lambda_L(x)$.  Our main focus in this paper will be classical and quantum gravitational effects in AdS$_3$, computed using recent work on Virasoro conformal blocks \cite{Fitzpatrick:2015dlt}.

\section{Quantum Corrections to Chaos from the Virasoro Identity Block}   
\label{sec:QuantumCorrections}

We are interested in studying gravitational interactions in AdS and their impact on chaos.  We will begin by reviewing known results and then discuss corrections in $1/c$, where $c$ is the central charge of the CFT$_2$.  

Graviton states in AdS$_3$ are created by the CFT$_2$ stress tensor, which has a holomorphic mode expansion $T(z) = \sum_n z^{-2-n} L_n$ in terms of the Virasoro generators $L_n$.  Thus all multi-graviton states in AdS$_3$ lie within a single irreducible representation of the Virasoro algebra.  The holomorphic Virasoro vacuum block $\mathcal{V}(z)$ represents of the exchange of all of these states between $WW$ and $VV$ in the correlator $F$ of equation (\ref{eq:OutOfOrderCorr}).  The full vacuum block contribution is a product $\mathcal{V}(z) \mathcal{V}(\bar z)$ of holomorphic and anti-holmorphic blocks, which depend independently on the holomorphic and anti-holomorphic dimensions $h_W, h_V$ and $\bar h_W, \bar h_V$, and on $c$.

The Virasoro vacuum block can only be computed in closed form in certain limits.  At leading order in the large central charge $c \to \infty$ limit with fixed holomorphic conformal dimensions $h_W$ and $h_V$, the Virasoro vacuum block reduces to the vacuum contribution plus a 1-graviton global conformal block
\be
\mathcal{V}(z) & = & 1 + \frac{2 h_W h_V}{c} z^2 {}_2 F_1(2,2,4,z) + O \left( \frac{1}{c^2} \right)
\nn \\
&=& 1 + \frac{2 h_W h_V}{c}  \left( \frac{6 (z-2) \log (1-z)}{z}-12 \right) + O \left( \frac{1}{c^2} \right)
\ee
This one-graviton contribution is sufficient to extract the Lyapunov exponent $\lambda_L$ at leading order in a $1/c$ expansion.  We see explicitly that the analytic continuation of Figure \ref{fig:AnalyticContinuationChaos} shifts $\log(1-z) \to \log(1-z) - 2 \pi i$.\footnote{One can pass to the second sheet by taking $(1-z) \rightarrow e^{2\pi i} (1-z)$ or $(1-z)\rightarrow e^{-2\pi i}(1-z)$.  In this subsection, we choose the latter in order to be consistent with the conventions in \cite{Roberts:2014ifa}.} 
Expanding the result at large $t$, or small $z = -4i e^{\frac{2\pi (x-t)}{\beta}}$, we have
\be \label{eq:1GravitonResult}
\mathcal{V}(z) \approx 1 + \frac{48 \pi i h_W h_V}{c z} = 1 - \frac{12 \pi h_W h_V}{c} e^{\frac{2 \pi}{\beta} (t-x)} 
\ee
This takes the form anticipated in equation (\ref{eq:Fapprox}), with $\lambda_L = \frac{2 \pi}{\beta}$.  Note that $\bar z \approx e^{-\frac{2 \pi}{\beta} (x+t)} \to 0$ in the relevant limit, and $\bar z$ has not been analytically continued off of the first sheet, so the anti-holomorphic $\mathcal{V}(\bar z) \approx 1$.

We are interested in the regime $\beta \ll t \ll t_*$, as depicted in Figure \ref{fig:ChaoticBehaviorLyapunov}.  Thus we can employ a slightly more systematic parameterization and fix $y = cz$ in the limit of large $c$.\footnote{In fact one could fix $c^r z$ with $0 < r < 1$, which would correspond to $t \propto r t_*$.  Fixing $z$  corresponds to taking $t$ completely independent of $c$ as $c \to \infty$, which in practice would mean $t \gtrsim t_d$.}  This does not alter equation (\ref{eq:1GravitonResult}), but it is useful in some more complicated examples.  

The Virasoro vacuum block can also be computed in the limit of large $c$ while fixing $h_W/c$ and $h_V$ to any value.  In this limit the $W$ operator corresponds to an object in AdS$_3$ with mass proportional to the Planck scale, while $V$ is a light probe with mass much less than the Planck scale.  In this heavy-light limit, the leading in $c$ \cite{Fitzpatrick:2015zha} contributions have been computed and matched to AdS$_3$ calculations, and more recently the sub-leading $1/c$ corrections have been computed  \cite{Fitzpatrick:2015dlt}.  After analytically continuing and expanding to leading order in $1/c$ with $y=cz$ fixed, one finds \cite{Roberts:2014ifa}
\be \label{eq:HeavyLightLeadingOrder}
F_\beta(t,x) \approx \left( \frac{1}{1 - \frac{24 \pi i h_W}{y} } \right)^{2 h_V} 
\ee
where $y \approx -4i e^{\frac{2\pi (x+t_*-t)}{\beta}}$ as follows from equation (\ref{eq:Sigmaoft}).
If we expand this result to first order in $1/c$ with fixed $z$ we match equation (\ref{eq:1GravitonResult}).  Thus the complete heavy-light Virasoro block does not provide any new information about the Lyapunov exponent as compared to 1-graviton exchange.  However, it does provide a nice case study for $F_\beta(t)$ \cite{Roberts:2014ifa}, as it displays the expected behavior for all times, pictured in Figure \ref{fig:ChaoticBehaviorLyapunov}.  In particular, in a $1/c$ expansion there are an infinite number of $1/(c z)^n = 1/y^n$ terms that individually have singular behavior at $y\sim 0$, and (\ref{eq:HeavyLightLeadingOrder}) is an explicit example of these resumming into something regular that actually vanishes at $y\rightarrow 0$. 

\begin{figure}[t!]
\begin{center}
\includegraphics[width=0.8\textwidth]{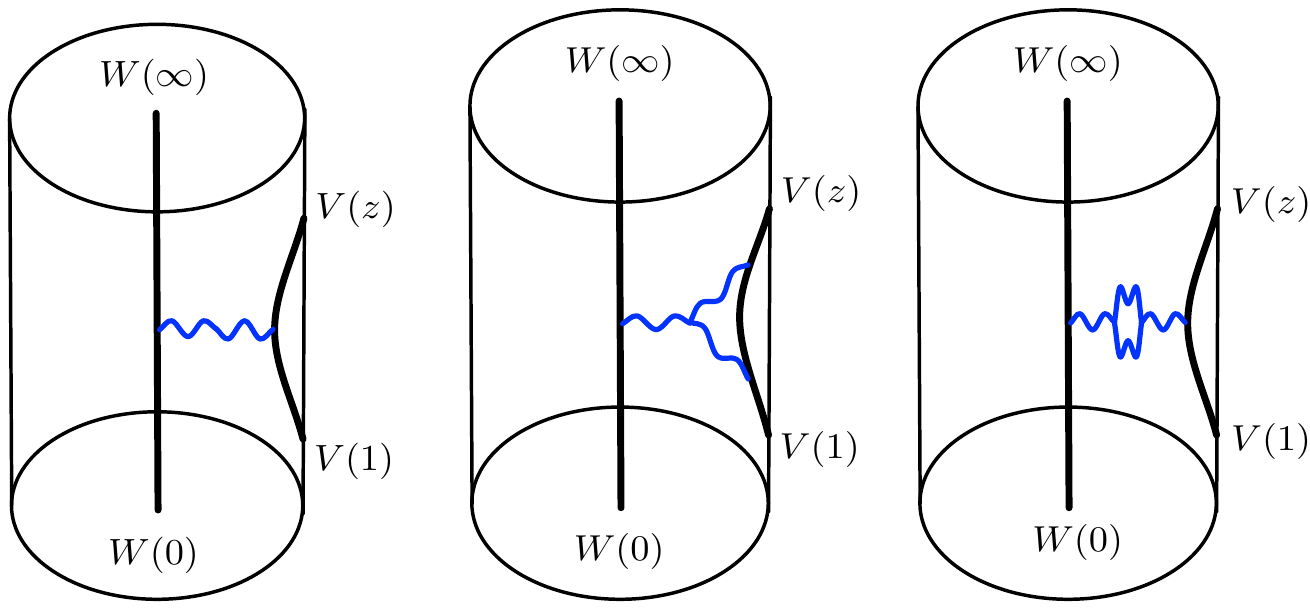}
\caption{ This figure shows diagrams in AdS$_3$ corresponding to various contributions \cite{Fitzpatrick:2015dlt} to the Virasoro vacuum block.  The diagram at left is 1-graviton exchange, and is proportional to $h_W h_V/c$.  The central diagram, proportional to $h_W h_V^2/c^2$ is a semi-classical correction incorporating gravitational back-reaction.  The diagram at right is a true quantum correction proportional to $h_W h_V/c^2$ which is responsible for $1/c$ corrections to the Lyapunov exponent.}
 \label{fig:OneOverCCorrectionsAdS} 
\end{center}
\end{figure}

Now let us consider $1/c$ corrections to these results, some of which correspond to quantum effects in AdS$_3$, as pictured in Figure \ref{fig:OneOverCCorrectionsAdS}.  These have been computed for the full heavy-light Virasoro vacuum block \cite{Fitzpatrick:2015dlt}, although we will only be interested in the low-order expansion in $h_W$ and $h_V$.

In the Euclidean region, we find an expression of the form
\be
\label{eq:PerturbativeVirasoroBlock}
\mathcal{V}(z) &=& 1 + \left( \frac{2 h_W h_V}{c} z^2 {}_2 F_1(2,2,4,z) \right) + \frac{1}{2} \left( \frac{2 h_W h_V}{c} z^2 {}_2 F_1(2,2,4,z) \right)^2 + \cdots
\nn \\
&& + \frac{h_W h_V^2 + h_W^2 h_V}{c^2} f_{\mathrm{LO,SC}}(z)  + h_V \frac{h_V}{c} \frac{ h_W^2}{c^2} f_{\mathrm{NLO,SC}}(z) + h_V \frac{1}{c} \frac{ h_W}{c} f_{\mathrm{Q}}(z) + \cdots
\ee
On the first line are the pure 1-graviton and 2-graviton contributions, neglecting the self-interactions of the gravitons.  The first term on the second line `$f_{\mathrm{LO,SC}}$' is so-labeled because it comes entirely from the leading order semi-classical heavy-light large $c$ limit.  In other words, it only involves information incorporated into equation (\ref{eq:HeavyLightLeadingOrder}).\footnote{The ``semi-classical limit'' is defined as the terms that survive in $\lim_{c\rightarrow \infty} \frac{1}{c} \log(\CV)$ when the ratios  $\eta_i \equiv h_i/c$ of the external operators to the central charge are all held fixed.  The `heavy-light' expansion of this semi-classical limit is an expansion in small $h_V/c$, with $h_W/c$ arbitrary.}  

In contrast, the term $f_{\mathrm{NLO,SC}}$ in equation (\ref{eq:PerturbativeVirasoroBlock}) is a next-to-leading-order correction to the semi-classical heavy-light block incorporating gravitational backreaction from the light probe $V$.  The third term $f_{\mathrm{Q}}$ corresponds to a true quantum correction in AdS$_3$, which will be responsible for a $1/c$ correction to the Lyapunov exponent.  Note that its coefficient translates into the AdS$_3$ expression
\be
\frac{h_V h_W}{c^2} \propto  \left( \frac{G_N}{R_{AdS}}  \right) G_N m_V m_W
\ee
where we have restored units, including the AdS curvature scale $R_{AdS}$.\footnote{Dimensional analysis is facilitated by studying $(G_N R_{AdS}) \log \mathcal{V}$, a dimensionless function.}  Thus the quantum correction should be viewed as an effect that disappears in the flat space limit where $R_{AdS} \to \infty$ with other parameters held fixed.  

We recently computed $f_{\mathrm{SC}}$ and $f_{\mathrm{Q}}$ explicitly \cite{Fitzpatrick:2015dlt}.  For the latter we find
\begin{equation}
\begin{aligned}
f_{\mathrm{Q}} &= -\frac{12}{z^2}  \left(-6 (z-2) z
   \left(\text{Li}_2\left(\frac{1}{1-z}\right)+\text{Li}_2(z)\right)+\left(\pi
   ^2 (z-2)-16 z\right) z\right. 
\\
   & \left. -3 (3 (z-2) z+2) \log ^2(1-z)+(z-2) z (6 \log (z)+6 i
   \pi -1) \log (1-z)\right) 
   \end{aligned}
   \label{eq:CorrectionsToV}
\end{equation} 

Performing the analytic continuation of Figure \ref{fig:AnalyticContinuationChaos} and expanding in $1/c$ with $y = cz$ fixed, we find 
\be \label{eq:ContinuedQuantumV}
\mathcal{V}(z) \approx 1 + 48 \pi i h_W h_V \left( \frac{1}{y} - \frac{12 \log \left(y / c\right)}{cy} + \frac{12 i \pi +7}{c y}   - \frac{3 i\log(y) }{\pi c^2} + \frac{6 i \pi}{y^2} \right) 
\ee
Now we will discuss each of the terms in parentheses.
The first term corresponds to the one-graviton exchange that we studied in equation (\ref{eq:1GravitonResult}), while the others are quantum corrections.  The third term is a complex $1/c$ correction to the overall coefficient of $e^{\lambda_L t}$.    The fourth term represents highly suppressed linear growth in $t$  not of Lyapunov form.

We would like to focus on the first two terms in parentheses in equation (\ref{eq:ContinuedQuantumV}).  It is natural to interpret these as arising from an expansion 
\be
\frac{1}{c \, z^{1 + \frac{12}{c}}} \approx \frac{1}{y} \left( 1 - \frac{12 \log \left(y /c \right)}{c} + \frac{1}{2} \left( \frac{12 \log \left(y /c \right)}{c} \right)^2 + \cdots \right)
\ee
which we would expect for the correlators of a CFT.
Recalling that $z = -4i e^{\frac{2\pi (x-t)}{\beta}}$, the logarthmic correction we have found can be interpreted as a positive quantum contribution to the Lyapunov exponent.     We have checked the resummation of these logarithms using a forthcoming \cite{Hongbin} computation of the $1/c^3$ corrections to the Virasoro vacuum block, verifying the coefficient of the $\log^2(y) / c^2$ term above. Furthermore, we have also found a general proof of leading logarithmic resummation \cite{Hongbin} to all orders in $\log(z)/c$.  Thus at next-to-leading order in $1/c$, by one reasonable definition the Lyapunov exponent is
\be
\lambda_L = \frac{2 \pi }{\beta} \left( 1 + \frac{12}{c} \right)
\ee
Naively interpreted, this violates the bound on chaos  \cite{Maldacena:2015waa}, but as we explained in section \ref{sec:ReviewBound}, the detailed analysis \cite{Maldacena:2015waa} permits $1/c$ corrections of either sign.  This effect might also be interpreted as a quantum shift in the graviton Regge intercept in AdS$_3$ \cite{JoaoRegge}, which would vanish in the flat spacetime limit.

However, before we conclude we must return to discuss the last term in parentheses in equation (\ref{eq:ContinuedQuantumV}).  Note that it has a relative phase of $i$ compared to the leading one-graviton term.  Thus it is tempting to interpret it as a quantum two-graviton correction akin to the last term on the first line of equation (\ref{eq:PerturbativeVirasoroBlock}), differentiating it from the Lyapunov exponent.  
Nevertheless, this last term contributes significantly to $F_\beta(t)$ in the relevant regime $\beta \ll t \ll t_*$, leading to 
\be \label{eq:TheQuantumCorrection}
F_\beta(t) \approx 1 - \frac{12 \pi h_W h_V}{c} e^{\frac{2 \pi }{\beta} \left( 1 + \frac{12}{c} \right) t} + \frac{18 \pi h_W h_V}{c^2} e^{\frac{4 \pi }{\beta}  t} + \cdots
\ee 
where we have neglected the third and fourth terms in parentheses in equation (\ref{eq:ContinuedQuantumV}).
If we view $\lambda_L$ as the first exponent, then it has received a $1/c$ correction, but this effect does not dominate over the other contributions to $F_\beta(t)$ in the physical regime.  We could also define $\lambda_L$ via $\log(1-F_\beta)$, but this would produce a $t$-dependent Lyapunov exponent.  Clearly $\mathcal{V}(z)$ and $F_\beta(t)$ are unambiguous functions, but depending on the definition we may interpret our result as either a constant $1/c$ correction to $\lambda_L$, a time-dependent $\lambda_L$, or simply as the failure of $\lambda_L$ to precisely capture the physics of scrambling at finite $c$.

\section{Local AdS Correlators, Scattering, and Causality}
\label{sec:AdS}

In this section we discuss several issues related to AdS correlators, bulk locality, and the connection between scrambling and scattering.  In section \ref{sec:DoubleTraceContributions} we study a full AdS$_3$ correlator, which includes both the vacuum Virasoro block and certain double-trace operator contributions.  In section \ref{sec:BulkPoint} we compare chaos and bulk scattering, noting how a certain `bulk point singularity' might be used to diagnose bulk locality near black hole horizons.  In section \ref{sec:CausalityAnalyticity} we connect our results on chaos to causality and analyticity in both AdS/CFT and flat spacetime scattering.

\subsection{Bulk Correlators and Double-Trace Operators}
\label{sec:DoubleTraceContributions}

In the previous section we used the Virasoro vacuum conformal block, which includes the exchange of all graviton states, to compute a quantum correction to the Lyapunov exponent in AdS$_3$/CFT$_2$.  However, even in case of pure gravity in AdS$_3$, the $\<WWVV\>$ correlator receives other contributions. One way of understanding this is that even a ``pure'' gravity theory (whose only low-energy degrees of freedom are gravitons) is not really pure once we introduce the probe operators $V,W$.  At a minimum, at large $c$, the theory contains multi-trace operators, corresponding to multi-particle states made from the probes $V$ and $W$.  From the point of view of the conformal block decomposition or the OPE, double-trace operators always make an important contribution to any perturbative process in AdS \cite{Liu}.  As we discussed in section \ref{sec:ReviewBound}, the crucial limit that suppresses these other contributions is $x \gg 1$ (i.e. $\bar{z} \ll 1$) where the zero-twist sector dominates.  In this section, we will consider an example where we can explicitly explore the size of double-trace contributions using a bulk computation \cite{KeskiVakkuri:1998nw}.  

Let us first consider the relatively simple case where the probe operators interact through the backreaction of $W$ on the bulk geometry.   Since the resulting geometry is a quotient of pure AdS$_3$, the correlator  for the CFT on a Lorentzian cylinder can be written as a sum over images \cite{KeskiVakkuri:1998nw}
\begin{equation}
\< V W V W\> =|1-z|^{-2h_V} \sum_{n=-\infty}^\infty \left( \frac{ \alpha^2}{4\sinh(\frac{\alpha}{2}( \log(1-z)+i  (n+2) \pi)) \sinh(\frac{\alpha}{2} (\log(1-\bar{z})-i n \pi))} \right)^{2h_V},
\label{eq:bulkcorr}
\end{equation}
where $\alpha^2 \equiv 1 - \frac{24 h_W}{c}$ and the $n+2$ arises because of the analytic continuation of $z$ to the Lorentzian sheet as depicted in Figure \ref{fig:AnalyticContinuationChaos}.

In the limit $\bar{z}\rightarrow 0$ and generic $\alpha$, the dominant contribution comes from the $n=0$ term, since this is the only one with an OPE singularity $\propto \bar{z}^{-2h_L}$.  Since this term is exactly the contribution from the large $c$ vacuum Virasoro block \cite{Fitzpatrick:2014vua, Fitzpatrick:2015zha, Fitzpatrick:2015foa, Asplund:2014coa, Hijano:2015rla, Hijano:2015qja}, we immediately see that contributions from multi-trace operators made from the probes $V$ and $W$ are irrelevant in this limit.  It will be useful nevertheless to warm up by being even more explicit here.   The $n=0$ term decays at very small $z$ relative to the disconnected correlator:
\be
\frac{\< V W V W\>}{\< VV\> \<WW\> } &\sim& \left( \frac{\alpha z}{2 \sin(\pi \alpha)} \right)^{2h_V}.
\ee
As usual, we are interested in the initial onset of this decay at large $c$.  Taking $\alpha \approx 1- \frac{12 h_W}{c}$, one obtains
\be
\frac{\< V W V W\>}{\<VV\> \<WW\>} &\approx& 1 - \frac{48 i  \pi h_V h_W}{z c} + \dots .
\ee
The first term at large $c$ is shown above, and agrees with equation (\ref{eq:1GravitonResult}).  To quantify the corrections, we take the limit $c\gg 1$ with $y=z c$ fixed; in this limit, the first few contributions are
\be
\frac{\< V W V W\>}{\<VV\> \<WW\>} &\approx& \left( 1 - \frac{48 i \pi h_V h_W}{y} + \CO(y^{-2}))\right) + \frac{1}{c} \left( 24i \pi h_V h_W + \CO(y^{-1}) \right) + \dots .
\ee
In fact, all terms that we have written out explicitly above come from just the single stress tensor contribution $\propto \frac{h_V h_W}{c}z^2 {}_2F_1(2,2,4,z)$. 

Now let us look at the bulk result in a more general regime, where $c$ is large with $y= z c$ and $\bar{y} = \bar{z} c$ fixed.  In this limit, after dividing by the disconnected correlators, equation (\ref{eq:bulkcorr})  becomes
\be
\frac{\< V W V W\> }{\< VV\>\<WW\>} &\approx& \sum_{n=-\infty \atop n ~ \textrm{even}}^\infty \left( \frac{1}{\left( 1 + \frac{12 i \pi h_W (n+2)}{y}\right)\left( 1 + \frac{12 i \pi h_W n}{\bar{y}} \right) } \right)^{2h_V}
\ee
Taking $h_V =1 $, the sum can be computed in closed form to give
\be
\frac{\< V W V W\> }{\< VV\>\<WW\>} &\approx&
\frac{y^2 \bar y^2  \left(\text{csch}^2\left(\frac{y}{12
   h_W}\right)+\text{csch}^2\left(\frac{\bar y}
   {12 h_W}\right)\right)}{144 h_W^2 \left(24 \pi i  h_W+ (y-\bar y)\right)^2}
\nn \\ 
&& + \frac{ y^2 \bar y^2 \left(  \coth
   \left(\frac{y}{12 h_W}\right)- \coth \left(\frac{\bar y}{12  h_W}\right)\right)}{6 h_W \left(24 \pi i h_W+ (y-\bar y)\right)^3}
\label{eq:BulkCorrelator}   
\ee 
The correlator will depend on both $x$ and the time $t$ discussed in section \ref{sec:ReviewBound} due to the contributions of double trace $V \partial^k V$ and $W \partial^k W$ type operators.  To see this explicitly, one can expand equation (\ref{eq:BulkCorrelator}) in $\bar y$, giving the vacuum Virasoro block as the leading term and a correction proportional to $\bar y^2$ which is associated to $V \partial^k V$ double trace operators in the conformal block expansion.

\begin{figure}[t!]
\begin{center}
\includegraphics[width=0.75\textwidth]{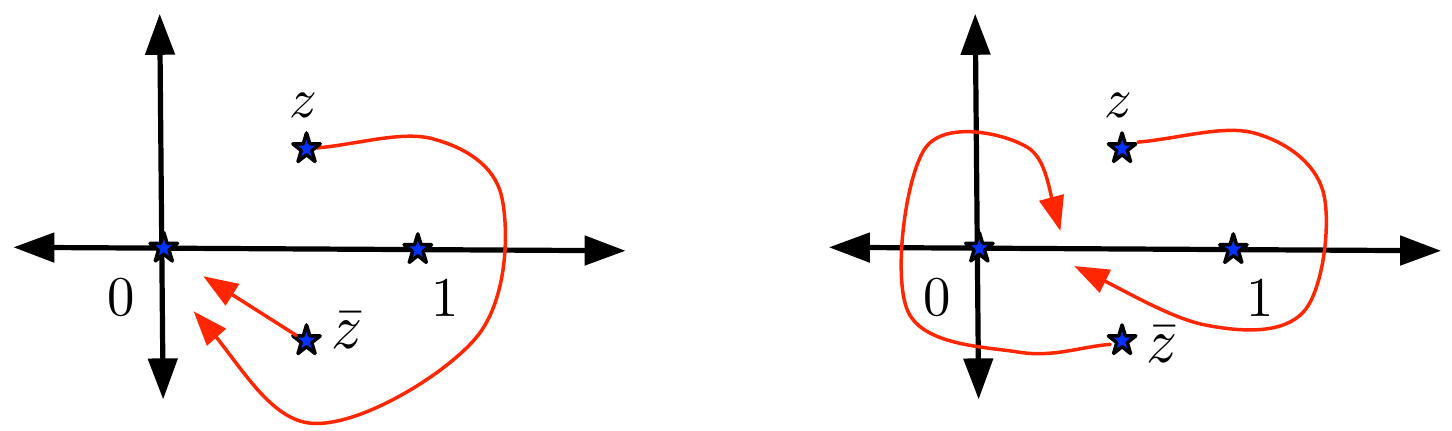}
\caption{ We compare the analytic continuation relevant for chaos (left) \cite{Roberts:2014ifa}, which requires crossing a single light cone branch cut, with the analytic continuation for the study of bulk scattering (right) \cite{GGP}, which requires crossing two branch cuts from the Euclidean region. }
 \label{fig:ContinuationsChaosvsScattering} 
\end{center}
\end{figure}

\subsection{Bulk Point Singularities and Scattering Near Horizons}
\label{sec:BulkPoint}

To address the black hole information paradox, it would be useful to be able  to use CFT data to define AdS observables  behind the horizon of a black hole.  However, not only is this problem difficult, but because we do not expect bulk observables to have a precise existence, it may not even be well-defined.

Fortunately, CFT correlators have some features with a precise bulk interpretation.  A particularly sharp example is the bulk point singularity \cite{GGP, JP, Okuda:2010ym, JoaoMellin, Maldacena:2015iua} of four-point correlators at $z = \bar z$, which is associated with AdS scattering amplitudes.  The singularity arises when a set of null rays emanating into the bulk from CFT operator insertions all meet at a bulk point. This occurs when $\det x_{ij}^2 = 16 (\rho - \bar \rho)^2(1-\rho \bar \rho)^2 \propto (z-\bar z)^2$ vanishes \cite{GGP}, where the $\rho$ coordinates are pictured in figure \ref{fig:OpLocation}.  Thus the singularity is a signature of the existence of a local bulk.  CFT correlators computed from AdS perturbation theory (Witten diagrams) generically have such bulk point singularities.  In CFT$_2$ these bulk point singularities are resolved at finite $c$ \cite{Maldacena:2015iua}, although the correlators may still grow a very large `bump' in the vicinity of $z=\bar z$.  The Lorentzian correlator of equation (\ref{eq:BulkCorrelator}) has such a bump at large $c$ with fixed $z, \bar z$, although it has an exponentially sensitive coefficient.
 
Bulk point singularities can never correspond with points behind the horizon of an AdS black hole.  This follows from the definition of a horizon -- null rays extending from behind the horizon will never reach the boundary of AdS.   But it is natural to ask if there exist bulk point singularities associated with bulk points very close to the horizon of a black hole.  In other words, do bulk point singularities `fuzz out' in a continuous way as we try to use them to probe local physics closer and closer to the horizon of a black hole?   
 
These considerations suggest a setup closely related to chaos and scrambling.  In AdS/CFT, scrambling arises from interactions near the horizon of a black hole \cite{Shenker:2013pqa, JoeRuelle}.  A crucial role is played by the universal blue-shift experienced by all infalling objects.  For example, consider a particle falling towards the horizon of a BTZ black hole with temperature $T$.  If the particle is released at a time $-t$, it will blueshift by a factor of $\sim \frac{1}{T} e^{t T}$ once it crosses the $t=0$ time slice \cite{Shenker:2013pqa}.  This relative blue-shift will also be important for scattering processes that occur in `the zone' near the horizon.  This means that in AdS$_3$/CFT$_2$ we can study scattering in a black hole background via a conformal transformation from the plane to the thermal cylinder.  The conformal transformation cannot create or eliminate bulk point singularities, but it does alter the kinematical interpretation of the region $z \sim \bar z$.  In specific theories with a bulk point `bump' \cite{Maldacena:2015iua} it may be possible to precisely characterize the limitations of bulk locality near horizons.  Correlators with more than four operators could be used to study bulk points in the background of perturbations that shift the location of the horizon \cite{Leichenauer:2014nxa, Roberts:2014isa, Shenker:2013yza}.

As we briefly review in appendix \ref{app:AnalyticContinuations} (see also \cite{Maldacena:2015iua}), the kinematics of bulk point singularities differs subtly from that of the correlators we have used to diagnose chaos.  However, the regimes overlap near $\bar z \sim 0$, the lightcone OPE limit.  This means that we can relate scattering and Lyapunov exponents more generally.
Let us consider introducing additional interactions in the bulk in a low-energy effective theory.  
For scalar fields $\phi_V, \phi_W$ in the bulk dual to $V,W$ on the boundary, a local quartic interaction with $2k$ derivatives, i.e. of the form
\be
(\partial_{\mu_1} \dots \partial_{\mu_s} \phi_V) (\partial^{\mu_1} \dots \partial^{\mu_s}\phi_V) (\partial_{\mu_{s+1}} \dots \partial_{\mu_k} \phi_W)(\partial^{\mu_{s+1}} \dots \partial^{\mu_k} \phi_W)
\ee
creates a leading bulk singularity (after the analytic continuation of Figure \ref{fig:ContinuationsChaosvsScattering}) at $z \sim \bar{z}$ of the form
\be
\frac{F_\beta}{\< VV\>\<WW\>} &\sim& \frac{1}{(z-\bar{z})^{\gamma}} z^{2\Delta_V+2\Delta_W+k-2} (1-z)^{\Delta_V + \Delta_W + k-2}\frac{T\left(-\frac{t}{s}= z\right)}{s^k}, \nn\\
\gamma &=& 2\Delta_V+2\Delta_W+2k-3,
\label{eq:bulksingform}
\ee
where $s^{-k}T\left(-\frac{t}{s}\right) \equiv T(\sin^2\frac{\theta}{2})$ is the angular dependence of the leading power of Mandelstam $s$ in the flat-space scattering amplitude.  This singularity overlaps with the region of large $x$ and large $t$ relevant for our study of chaos, where both $z$ and $\bar{z}$ approach zero.  Note that this is given by the forward limit $T(0)$ of the scattering amplitude; we will comment below in more detail on the connection with analyticity constraints on the forward scattering limit in flat-space \cite{Adams:2006sv}. At small $z$, this bulk singularity grows like\footnote{The order of limits here is not quite the same as the one that is relevant for the MSS bound, since we are taking $z\sim \bar{z}$ first and then taking $z\rightarrow 0$.  Thus the residue in (\ref{eq:bulksingform}) does not keep track of the difference between $z$ and $\bar{z}$ when these both approach zero but at different rates. In particular, in the limit where both approach zero with a large ratio $\eta = \bar{z}/z$, the final result can contain powers of $\eta$, and we are not keeping track of these on the RHS of (\ref{eq:bulksingMSS}). }
\be
\frac{F_\beta}{\< VV\>\<WW\>} &\sim& \frac{ T(0)}{z^{k-1}}.
\label{eq:bulksingMSS}
\ee
Therefore this violates the MSS bound \cite{Maldacena:2015waa} on the rate of growth for $k>2$, and for $k=2$ it is marginal.  In particular, when $k=2$, any real $T(0)$ must have the correct sign so that this contribution to the disconnected piece {\it decreases} its magnitude, since an increase would violate (\ref{eq:boundsummary}). This observation was one of the central results of a recent study of causality in CFT \cite{Hartman:2015lfa}.  The fact that $k>2$ violates the bound implies that such bulk interactions are much like higher-spin contributions to CFT correlators in the context of the MSS bound.  That is, each one individually violates the bound, and the only way for them to be present in a consistent unitary theory is via an infinite number of higher derivative interactions with correlated coefficients that resum and soften the singularity.\footnote{We thank Tom Hartman for emphasizing this to us.}   

\subsection{Causality and Analyticity in Flat Space vs AdS/CFT}
\label{sec:CausalityAnalyticity}

The above constraints bear a striking resemblance to certain bounds on effective field theories derived from the analyticity of flat-space scattering amplitudes \cite{Adams:2006sv}.  In fact, with a little extra work using results \cite{GGP} connecting the bulk point singularity to the flat space S-Matrix, one can see that the integration contour on the forward limit of the flat space scattering amplitude  \cite{Adams:2006sv} is in fact the same as the integration contour used  \cite{Hartman:2015lfa} to study causality in CFT. 

 Let us use the form (\ref{eq:bulksingform}) for the bulk singularity near $\bar{z} \sim 0$ and  $z= -t/s \ll 1 $  to write
\be
T(s,t \ll s) &\sim& s^k \left( -\frac{t}{s} \right)^{k-1} \left( -\frac{u}{s} \right)^{2+k-\Delta_V-\Delta_W} \left( \frac{z-\bar{z}}{z} \right)^{\gamma} G\left(- \frac{t}{s}\right) \nn\\
 &\sim& s t^{k-1} G\left( -\frac{t}{s}\right),
 \label{eq:TtoG}
\ee
where $G(z) \equiv \lim_{\bar{z} \rightarrow 0} \frac{F_\beta}{\< VV\>\<WW\>}$.  Now, consider the case of a bulk interaction $(\partial \phi)^4$ with $V=W, \phi_V = \phi_W$, which corresponds to $k=2$ above, and was one of the main cases of interest in \cite{Hartman:2015lfa} and \cite{Adams:2006sv}.  To derive analyticity contraints on the forward limit of scattering amplitudes one studies a contour integral (see figure 8 of \cite{Adams:2006sv}) 
\be
\lim_{t\rightarrow 0} \oint ds \frac{T(s,t)}{s^3}
\label{eq:NimaContour}
\ee
By isolating the contributions of the interaction $(\partial \phi)^4$ and using unitarity, one can  prove  \cite{Adams:2006sv} that this operator must have a positive coefficient in effective field theory.
If we perform a change of variables to $\sigma = -\frac{t}{s}$, this contour integral becomes
\be
\lim_{t\rightarrow 0} \oint ds \frac{T(s,t)}{s^3} \sim \oint d\sigma \, G(\sigma),
\ee
where we used equation (\ref{eq:TtoG}).  Note that the semi-circular $\sigma$ contour must be taken at small radius because $t/s$ is small.\footnote{See section 6.4 of \cite{Hartman:2015lfa} with $\ell_m=2$. }  This is the contour integral that was recently used (see figure 6 of \cite{Hartman:2015lfa}) to study causality constraints in CFT, and to prove e.g. that AdS effective theories must have positive coefficients for $(\partial \phi)^4$ interactions, though in the limit $\bar{z}/z \ll1$ first rather than $z-\bar{z}\ll 1$ first.   So there is a direct connection between causality constraints on flat space and AdS effective field theories.

\section{Discussion}
\label{sec:discussion}

We have argued that one-loop gravitational effects in AdS$_3$, which correspond to universal $1/c$ corrections to the vacuum Virasoro conformal block of any large $c$ CFT$_2$, produce a quantum correction to chaos encapsulated by equation (\ref{eq:TheQuantumCorrection}).  The result might be viewed as a correction to the Lyapunov exponent
\be
\lambda_L = \frac{2 \pi}{\beta} \left( 1 + \frac{12}{c} \right)
\ee
but the interpretation is somewhat ambiguous, since other $1/c$ effects are of similar or greater importance between the dissipation and scrambling times.   If interpreted as the $\lambda_L$ above, the result violates a recently proposed bound on chaos \cite{Maldacena:2015iua}, but it does not contradict the arguments that led to that bound, or its spirit \cite{Sekino:2008he} that black holes may be the fastest scramblers.  Viewed as a correction to gravitational scattering in AdS$_3$, the effect we have identified is proportional to $\frac{G_N}{R_{AdS}}$, so it is a long-distance effect that would vanish in the flat spacetime limit.

In section \ref{sec:Bound} we specialized the arguments for the bound \cite{Maldacena:2015iua} to CFT$_2$ in order to examine the potential for $1/c$ corrections.  We found that for $t \gtrsim t_d$ near the dissipation time corrections to the bound can be of order $1/c$, but that for times $t \sim t_*$ near the scrambling time, any positive corrections to the bound must be parametrically suppressed by $\propto 1/c^3$ in CFT$_2$.  In accord with these results, the corrections to $\lambda_L$  are only positive at very early times.

For more general systems the role of  $1/c$ will be played by $1/N^2$, or the inverse of the parameter controlling the number of degrees of freedom.  Thus our analysis raises the question of whether other effects violating the bound may be identified in other systems.  Optimistically, we may hope that pure gravity always produces the largest possible value of $\lambda_L$ once all $1/c$ corrections are taken into account.  Then we must ask whether this idea could be well-defined in higher dimensions.  In AdS$_{> 3}$ we  expect similar quantum corrections to $\lambda_L$ in the perturbative $G_N$ expansion, but if they come from AdS-scale effects, ie if they are parametrically of order $\frac{G_N}{R_{AdS}^{d-1}}$, then perhaps they will be relatively universal.  Thus the bound on $\lambda_L$ may remain more precise \cite{Roberts:2014isa} than the KSS bound \cite{Kovtun:2004de} on the viscosity/entropy ratio, which can be violated \cite{Kats:2007mq, Brigante:2007nu} by higher dimension local operators in the gravitational action.  A more pessimistic interpretation would be that the bound on $\lambda_L$ cannot be made completely sharp at finite but large values of $N$, and only emerges in the strict $N \to \infty$ limit.   Even in this case, the bound should still apply to effects (e.g. stringy corrections) that depend on a parameter (e.g. the 't Hooft coupling) that can be adjusted independently in the large $N$ limit.

We have also discussed the relationship between chaos, scattering, AdS locality, and causality constraints on CFT correlators and flat space scattering amplitudes, emphasizing that many recent constraints are closely related \cite{Adams:2006sv, Camanho:2014apa,  Maldacena:2015iua, Hartman:2015lfa} and can be more directly connected.  We explained that the bulk point singularity, a signature of local bulk scattering \cite{GGP, JP, JoaoMellin, Maldacena:2015iua}, may be used to examine bulk locality near horizons via a kinematical setup \cite{Shenker:2013pqa} closely related to scrambling.

\section*{Acknowledgments}

We would like to thank Brian Swingle for discussions and collaboration at early stages of this work, and Ethan Dyer  and Dan Roberts for comments on the draft.  We would also like to thank Hongbin Chen, Ethan Dyer, Tom Hartman, Ami Katz, Daliang Li, Eric Perlmutter, Dan Roberts,  Steve Shenker, Matt Walters, Junpu Wang, and Sasha Zhiboedov for valuable discussions.  JK  is supported in part by NSF grants PHY-1316665 and PHY-1454083, and by a Sloan Foundation fellowship. 

\begin{appendices}

\section{Details of Analytic Continuations}
\label{app:AnalyticContinuations}

Let us first discuss the analytic continuation that takes the correlator from the Euclidean sheet to the second sheet relevant to chaos and to the Regge limit.
 Under the analytic continuation of Figure \ref{fig:AnalyticContinuationChaos} the various logarithms and polylogarithms have monodromies
\be 
\log(1-z) &\to& \log(1-z) - 2 \pi i,
\nn \\
\mathrm{Li}_n(z) &\to & \mathrm{Li}_n(z) + \frac{2 \pi i }{(n-1)!} \log^{n-1}(z), \nn\\
\mathrm{Li}_n(1-z) & \to & \mathrm{Li}_n(1-z) ,
\ee
which can be easily derived from $\mathrm{Li}_n(z) = \int_0^z \frac{\mathrm{Li}_{n-1}(t)}{t} dt$ and $\mathrm{Li}_1(z) = -\log(1-z)$.

The analytic continuation above differs from that required for an analysis of scattering amplitudes (in the flat spacetime limit) and the bulk point singularity  \cite{GGP, JP, Maldacena:2015iua}.  The latter require continuation through a second branch cut and onto a further sheet, as depicted in Figure \ref{fig:ContinuationsChaosvsScattering}.  
As a very explicit example illustrating the distinction, perturbative $\lambda \phi^4$ contact interactions in AdS produce correlators proportional to `$D$-functions'.  The $D_{1111}$ function has the closed form expression \cite{GGP} 
\be
D_{1111}^{\mathrm{Euc}} = \frac{z \bar z}{z - \bar z} \left(2 \mathrm{Li}_2(z) - 2\mathrm{Li}_2 (\bar z) +  \log( z\bar z) \log \left( \frac{1-z}{1 - \bar z} \right) \right)
\ee
on the Euclidean sheet.  Following Figure \ref{fig:ContinuationsChaosvsScattering}, on the sheet relevant for chaos the function continues to
\be
D_{1111}^{\mathrm{Chaos}} = \frac{z \bar z}{z - \bar z} \left( 2\mathrm{Li}_2(z) - 2 \mathrm{Li}_2 (\bar z) + 4 \pi i \log(z) + \log(z \bar z) \left( \log \left( \frac{1-z}{1 - \bar z} \right)  - 2 \pi i \right) \right)
\ee
whereas after the analytic continuation to the sheet relevant for scattering \cite{GGP, JP, Maldacena:2015iua} we have
\be
D_{1111}^{\mathrm{Scatter}} = \frac{z \bar z}{z - \bar z} \left( 2\mathrm{Li}_2(z) - 2 \mathrm{Li}_2 (\bar z)  + 4 \pi i \log(z) + (\log(z \bar z)+ 2 \pi i)\left( \log \left( \frac{1-z}{1 - \bar z} \right)  - 2 \pi i \right) \right)
\nn
\ee
The continuation relevant for scattering has a bulk point singularity at $z=\bar z$ \cite{GGP, JP, Maldacena:2015iua}, whereas $D_{1111}^{\mathrm{Euc}}$ and $D_{1111}^{\mathrm{Chaos}}$ are regular at these points.  This can be seen explicitly by expanding the functions above in the small parameter $z - \bar z$.  A bulk point singularity can only arise from a sum over an infinite number of conformal blocks, which makes the distinction between the two analytic continuations in figure \ref{fig:ContinuationsChaosvsScattering} somewhat subtle in the context of the conformal block decomposition.  For a more detailed discussion see section 6 of \cite{Maldacena:2015iua}.

\end{appendices}

\newpage

\bibliographystyle{utphys}
\bibliography{VirasoroBib}

\begin{thebibliography}{10}
\ifx\href\asklfhas\newcommand{\href}[2]{#2}\fi
\ifx\arxivref\asklfhas\newcommand{\arxivref}[2]{\href{http://arxiv.org/abs/#1}{#2}}\fi
\ifx\doiref\asklfhas\newcommand{\doiref}[2]{\href{http://dx.doi.org/#1}{#2}}\fi
\parskip 0pt
\normalsize

\bibitem{georges2001quantum}
A.~Georges, O.~Parcollet \& S.~Sachdev,
\textit{``Quantum fluctuations of a nearly critical Heisenberg spin glass''},
Physical~Review~B \textbf{63}, 134406 (2001).

\bibitem{d2015quantum}
L.~D'Alessio, Y.~Kafri, A.~Polkovnikov \& M.~Rigol,
\textit{``From Quantum Chaos and Eigenstate Thermalization to Statistical
  Mechanics and Thermodynamics''},
\normalsize{\texttt{\arxivref{1509.06411}{arXiv:1509.06411}}}.

\bibitem{bradbury1952sound}
R.~Bradbury,
\textit{``{A Sound of Thunder}''},
Collier's \textbf{June 28}, 20 (1952).

\bibitem{kitaev2014hidden}
A.~Kitaev,
\textit{``Hidden correlations in the hawking radiation and thermal noise''},
in \textit{``talk given at Fundamental Physics Prize Symposium''}.
\bibitem{Shenker:2013pqa}
S.~H. Shenker \& D.~Stanford,
\textit{``{Black holes and the butterfly effect}''},
\doiref{10.1007/JHEP03(2014)067}{JHEP \textbf{1403}, 067 (2014)},
\normalsize{\texttt{\arxivref{1306.0622}{arXiv:1306.0622}}}.

\bibitem{Hosur:2015ylk}
P.~Hosur, X.-L. Qi, D.~A. Roberts \& B.~Yoshida,
\textit{``{Chaos in quantum channels}''},
\normalsize{\texttt{\arxivref{1511.04021}{arXiv:1511.04021}}}.

\bibitem{Gur-Ari:2015rcq}
G.~Gur-Ari, M.~Hanada \& S.~H. Shenker,
\textit{``{Chaos in Classical D0-Brane Mechanics}''},
\normalsize{\texttt{\arxivref{1512.00019}{arXiv:1512.00019}}}.

\bibitem{Berenstein:2015yxu}
D.~Berenstein \& A.~M. Garcia-Garcia,
\textit{``{Universal quantum constraints on the butterfly effect}''},
\normalsize{\texttt{\arxivref{1510.08870}{arXiv:1510.08870}}}.

\bibitem{Stanford:2015owe}
D.~Stanford,
\textit{``{Many-body chaos at weak coupling}''},
\normalsize{\texttt{\arxivref{1512.07687}{arXiv:1512.07687}}}.

\bibitem{Maldacena:2015waa}
J.~Maldacena, S.~H. Shenker \& D.~Stanford,
\textit{``{A bound on chaos}''},
\normalsize{\texttt{\arxivref{1503.01409}{arXiv:1503.01409}}}.

\bibitem{Shenker:2014cwa}
S.~H. Shenker \& D.~Stanford,
\textit{``{Stringy effects in scrambling}''},
\doiref{10.1007/JHEP05(2015)132}{JHEP \textbf{1505}, 132 (2015)},
\normalsize{\texttt{\arxivref{1412.6087}{arXiv:1412.6087}}}.

\bibitem{Fitzpatrick:2014vua}
A.~L. Fitzpatrick, J.~Kaplan \& M.~T. Walters,
\textit{``{Universality of Long-Distance AdS Physics from the CFT
  Bootstrap}''},
\doiref{10.1007/JHEP08(2014)145}{JHEP \textbf{1408}, 145 (2014)},
\normalsize{\texttt{\arxivref{1403.6829}{arXiv:1403.6829}}}.

\bibitem{Fitzpatrick:2015zha}
A.~L. Fitzpatrick, J.~Kaplan \& M.~T. Walters,
\textit{``{Virasoro Conformal Blocks and Thermality from Classical Background
  Fields}''},
\normalsize{\texttt{\arxivref{1501.05315}{arXiv:1501.05315}}}.

\bibitem{Fitzpatrick:2015foa}
A.~L. Fitzpatrick, J.~Kaplan, M.~T. Walters \& J.~Wang,
\textit{``{Hawking from Catalan}''},
\normalsize{\texttt{\arxivref{1510.00014}{arXiv:1510.00014}}}.

\bibitem{Asplund:2014coa}
C.~T. Asplund, A.~Bernamonti, F.~Galli \& T.~Hartman,
\textit{``{Holographic Entanglement Entropy from 2d CFT: Heavy States and Local
  Quenches}''},
\doiref{10.1007/JHEP02(2015)171}{JHEP \textbf{1502}, 171 (2015)},
\normalsize{\texttt{\arxivref{1410.1392}{arXiv:1410.1392}}}.

\bibitem{Hijano:2015rla}
E.~Hijano, P.~Kraus \& R.~Snively,
\textit{``{Worldline approach to semi-classical conformal blocks}''},
\doiref{10.1007/JHEP07(2015)131}{JHEP \textbf{1507}, 131 (2015)},
\normalsize{\texttt{\arxivref{1501.02260}{arXiv:1501.02260}}}.

\bibitem{Alkalaev:2015wia}
K.~B. Alkalaev \& V.~A. Belavin,
\textit{``{Classical conformal blocks via AdS/CFT correspondence}''},
\doiref{10.1007/JHEP08(2015)049}{JHEP \textbf{1508}, 049 (2015)},
\normalsize{\texttt{\arxivref{1504.05943}{arXiv:1504.05943}}}.

\bibitem{Hijano:2015qja}
E.~Hijano, P.~Kraus, E.~Perlmutter \& R.~Snively,
\textit{``{Semiclassical Virasoro Blocks from AdS$_3$ Gravity}''},
\normalsize{\texttt{\arxivref{1508.04987}{arXiv:1508.04987}}}.

\bibitem{Fitzpatrick:2015dlt}
A.~L. Fitzpatrick \& J.~Kaplan,
\textit{``{Conformal Blocks Beyond the Semi-Classical Limit}''},
\normalsize{\texttt{\arxivref{1512.03052}{arXiv:1512.03052}}}.

\bibitem{Hongbin}
H.~Chen, A.~L. Fitzpatrick, J.~Kaplan, D.~Li \& J.~Wang,
\textit{``{Forthcoming}''}.

\bibitem{HartmanLargeC}
T.~Hartman,
\textit{``{Entanglement Entropy at Large Central Charge}''},
\normalsize{\texttt{\arxivref{1303.6955}{arXiv:1303.6955}}}.

\bibitem{Brown:1986nw}
J.~D. Brown \& M.~Henneaux,
\textit{``{Central Charges in the Canonical Realization of Asymptotic
  Symmetries: An Example from Three-Dimensional Gravity}''},
\doiref{10.1007/BF01211590}{Commun.Math.Phys. \textbf{104}, 207 (1986)}.

\bibitem{Headrick}
M.~Headrick,
\textit{``{Entanglement Renyi entropies in holographic theories}''},
\doiref{10.1103/PhysRevD.82.126010}{Phys.Rev. \textbf{D82}, 126010 (2010)},
\normalsize{\texttt{\arxivref{1006.0047}{arXiv:1006.0047}}}.

\bibitem{Roberts:2014ifa}
D.~A. Roberts \& D.~Stanford,
\textit{``{Two-dimensional conformal field theory and the butterfly effect}''},
\normalsize{\texttt{\arxivref{1412.5123}{arXiv:1412.5123}}}.

\bibitem{Hartman:2015lfa}
T.~Hartman, S.~Jain \& S.~Kundu,
\textit{``{Causality Constraints in Conformal Field Theory}''},
\normalsize{\texttt{\arxivref{1509.00014}{arXiv:1509.00014}}}.

\bibitem{Maldacena:2015iua}
J.~Maldacena, D.~Simmons-Duffin \& A.~Zhiboedov,
\textit{``{Looking for a bulk point}''},
\normalsize{\texttt{\arxivref{1509.03612}{arXiv:1509.03612}}}.

\bibitem{Pappadopulo:2012jk}
D.~Pappadopulo, S.~Rychkov, J.~Espin \& R.~Rattazzi,
\textit{``{OPE Convergence in Conformal Field Theory}''},
\normalsize{\texttt{\arxivref{1208.6449}{arXiv:1208.6449}}}.

\bibitem{Fitzpatrick:2012yx}
A.~L. Fitzpatrick, J.~Kaplan, D.~Poland \& D.~Simmons-Duffin,
\textit{``{The Analytic Bootstrap and AdS Superhorizon Locality}''},
\doiref{10.1007/JHEP12(2013)004}{JHEP \textbf{1312}, 004 (2013)},
\normalsize{\texttt{\arxivref{1212.3616}{arXiv:1212.3616}}}.

\bibitem{KomargodskiZhiboedov}
Z.~Komargodski \& A.~Zhiboedov,
\textit{``{Convexity and Liberation at Large Spin}''},
\doiref{10.1007/JHEP11(2013)140}{JHEP \textbf{1311}, 140 (2013)},
\normalsize{\texttt{\arxivref{1212.4103}{arXiv:1212.4103}}}.

\bibitem{GGP}
M.~Gary, S.~B. Giddings \& J.~Penedones,
\textit{``{Local bulk S-matrix elements and CFT singularities}''},
\doiref{10.1103/PhysRevD.80.085005}{Phys.~Rev. \textbf{D80}, 085005 (2009)},
\normalsize{\texttt{\arxivref{0903.4437}{arXiv:0903.4437}}}.

\bibitem{Polchinski:2002jw}
J.~Polchinski \& M.~J. Strassler,
\textit{``{Deep inelastic scattering and gauge / string duality}''},
\doiref{10.1088/1126-6708/2003/05/012}{JHEP \textbf{0305}, 012 (2003)},
\normalsize{\texttt{\arxivref{hep-th/0209211}{hep-th/0209211}}}.

\bibitem{JP}
I.~Heemskerk, J.~Penedones, J.~Polchinski \& J.~Sully,
\textit{``{Holography from Conformal Field Theory}''},
\doiref{10.1088/1126-6708/2009/10/079}{JHEP \textbf{0910}, 079 (2009)},
\normalsize{\texttt{\arxivref{0907.0151}{arXiv:0907.0151}}}.

\bibitem{JoaoRegge}
M.~S. Costa, V.~Goncalves \& J.~Penedones,
\textit{``{Conformal Regge theory}''},
\normalsize{\texttt{\arxivref{1209.4355}{arXiv:1209.4355}}}.

\bibitem{Liu}
H.~Liu,
\textit{``{Scattering in anti-de Sitter space and operator product
  expansion}''},
\doiref{10.1103/PhysRevD.60.106005}{Phys.~Rev. \textbf{D60}, 106005 (1999)},
\normalsize{\texttt{\arxivref{hep-th/9811152}{hep-th/9811152}}}.

\bibitem{KeskiVakkuri:1998nw}
E.~Keski-Vakkuri,
\textit{``{Bulk and boundary dynamics in BTZ black holes}''},
\doiref{10.1103/PhysRevD.59.104001}{Phys.~Rev. \textbf{D59}, 104001 (1999)},
\normalsize{\texttt{\arxivref{hep-th/9808037}{hep-th/9808037}}}.

\bibitem{Okuda:2010ym}
T.~Okuda \& J.~Penedones,
\textit{``{String scattering in flat space and a scaling limit of Yang-Mills
  correlators}''},
\doiref{10.1103/PhysRevD.83.086001}{Phys.~Rev. \textbf{D83}, 086001 (2011)},
\normalsize{\texttt{\arxivref{1002.2641}{arXiv:1002.2641}}}.

\bibitem{JoaoMellin}
J.~Penedones,
\textit{``{Writing CFT correlation functions as AdS scattering amplitudes}''},
\doiref{10.1007/JHEP03(2011)025}{JHEP \textbf{1103}, 025 (2011)},
\normalsize{\texttt{\arxivref{1011.1485}{arXiv:1011.1485}}}.

\bibitem{JoeRuelle}
J.~Polchinski,
\textit{``{Chaos in the black hole S-matrix}''},
\normalsize{\texttt{\arxivref{1505.08108}{arXiv:1505.08108}}}.

\bibitem{Leichenauer:2014nxa}
S.~Leichenauer,
\textit{``{Disrupting Entanglement of Black Holes}''},
\doiref{10.1103/PhysRevD.90.046009}{Phys.~Rev. \textbf{D90}, 046009 (2014)},
\normalsize{\texttt{\arxivref{1405.7365}{arXiv:1405.7365}}}.

\bibitem{Roberts:2014isa}
D.~A. Roberts, D.~Stanford \& L.~Susskind,
\textit{``{Localized shocks}''},
\doiref{10.1007/JHEP03(2015)051}{JHEP \textbf{1503}, 051 (2015)},
\normalsize{\texttt{\arxivref{1409.8180}{arXiv:1409.8180}}}.

\bibitem{Shenker:2013yza}
S.~H. Shenker \& D.~Stanford,
\textit{``{Multiple Shocks}''},
\doiref{10.1007/JHEP12(2014)046}{JHEP \textbf{1412}, 046 (2014)},
\normalsize{\texttt{\arxivref{1312.3296}{arXiv:1312.3296}}}.

\bibitem{Adams:2006sv}
A.~Adams, N.~Arkani-Hamed, S.~Dubovsky, A.~Nicolis \& R.~Rattazzi,
\textit{``{Causality, analyticity and an IR obstruction to UV completion}''},
\doiref{10.1088/1126-6708/2006/10/014}{JHEP \textbf{0610}, 014 (2006)},
\normalsize{\texttt{\arxivref{hep-th/0602178}{hep-th/0602178}}}.

\bibitem{Sekino:2008he}
Y.~Sekino \& L.~Susskind,
\textit{``{Fast Scramblers}''},
\doiref{10.1088/1126-6708/2008/10/065}{JHEP \textbf{0810}, 065 (2008)},
\normalsize{\texttt{\arxivref{0808.2096}{arXiv:0808.2096}}}.

\bibitem{Kovtun:2004de}
P.~Kovtun, D.~T. Son \& A.~O. Starinets,
\textit{``{Viscosity in strongly interacting quantum field theories from black
  hole physics}''},
\doiref{10.1103/PhysRevLett.94.111601}{Phys.~Rev.~Lett. \textbf{94}, 111601
  (2005)},
\normalsize{\texttt{\arxivref{hep-th/0405231}{hep-th/0405231}}}.

\bibitem{Kats:2007mq}
Y.~Kats \& P.~Petrov,
\textit{``{Effect of curvature squared corrections in AdS on the viscosity of
  the dual gauge theory}''},
\doiref{10.1088/1126-6708/2009/01/044}{JHEP \textbf{0901}, 044 (2009)},
\normalsize{\texttt{\arxivref{0712.0743}{arXiv:0712.0743}}}.

\bibitem{Brigante:2007nu}
M.~Brigante, H.~Liu, R.~C. Myers, S.~Shenker \& S.~Yaida,
\textit{``{Viscosity Bound Violation in Higher Derivative Gravity}''},
\doiref{10.1103/PhysRevD.77.126006}{Phys.~Rev. \textbf{D77}, 126006 (2008)},
\normalsize{\texttt{\arxivref{0712.0805}{arXiv:0712.0805}}}.

\bibitem{Camanho:2014apa}
X.~O. Camanho, J.~D. Edelstein, J.~Maldacena \& A.~Zhiboedov,
\textit{``{Causality Constraints on Corrections to the Graviton Three-Point
  Coupling}''},
\normalsize{\texttt{\arxivref{1407.5597}{arXiv:1407.5597}}}.

\end{thebibliography}

\end{document}